# High-power doughnut beam generation for atom-funnel experiment

S. M. Iftiquar[A], H. Ito[A,B], K. Totsuka[A], A. Takamizawa[B], and M. Ohtsu[A,B]

[A]ERATO Localized Photon Project, Japan Science and Technology Corporation, 687-1-17/4F Tsuruma, Machida-shi, Tokyo

[B]Interdisciplinary Graduate School of Science and Engineering, Tokyo Institute of Technology, 4259 Nagatsuta, Kanagawa

We have generated a doughnut beam suitable for atom funnel experiment with the conversion efficiency of about 50 %. Cold rubidium atoms are generated from a magneto-optical trap (MOT) and collected in a hollow glass prism below the MOT. In this near-field atom funnel experiment [1,2], the evanescent field is created at the inner wall with a high power laser beam. The free-falling cold Rb atoms bounce at the inner wall, cooled by the evanescent field and collected through a 200 micron hole at the bottom. For efficient funneling and de-excitation of the funneled atoms, a doughnut beam should have a power more than 500 mW with a dark center of around 250 micron and a ring diameter of 5 mm. Figure 1 shows the experimental layout. A cw Ti:sapphire laser beam (780 nm single mode) is devided into two beams with the help of beam splitter. Two convex lenses $L_1$, $L_2$ are used to convert the plane wave fronts into spherical and then the two beams are overlapped after a suitable relative phase delay through PZT actuators mounted at one of the reflecting mirrors. The lens $L_3$ collimates the resulting output beam. This system is suitable mainly because the output doughnut beam is of high power. Another advantage is that the beam shape is tunable from a very small dark center to a very wide dark zone [3]. Figure 2 shows a CCD image of the doughnut beam the characteristics of which is desirable for the atom funnel experiment. We have observed a continuous variation of the dark center from zero diameter, which resembles to Airy disk, to 1.2 mm wide central dark zone [3]. An experimental observation and analytical investigation reveal that such a tuning covers various stages of the Hermite-Gaussian (HG) and the Laguerre-Gaussian (LG) modes. The HG mode is thought to lead to narrow dark region and high output power, while for the LG mode the central dark region widens with a low output power. With an 800 mW input Gaussian laser beam, the power of the output LG mode is 350 mW and the reflected power is 350 mW, while for the HG mode the output power is 410 mW and the reflected power is 290 mW. The reflected power is the beam power that returns back towards the laser. At about 1W laser power the output HG mode is 525 mW and that of LG mode is 440 mW with the reflected beam power as 350 mW and 440 mW, respectively. Thus a conversion efficiency of more than 50 % is achievable with a smaller dark center. The Gouy phase shift plays a significant role in this experiment. While the two beams are astigmatically focussed with different lenses $L_1$ and $L_2$, an axial phase shift results. With such a phase redistribution when the two beams interfere, we obtain a tunable doughnut beam.

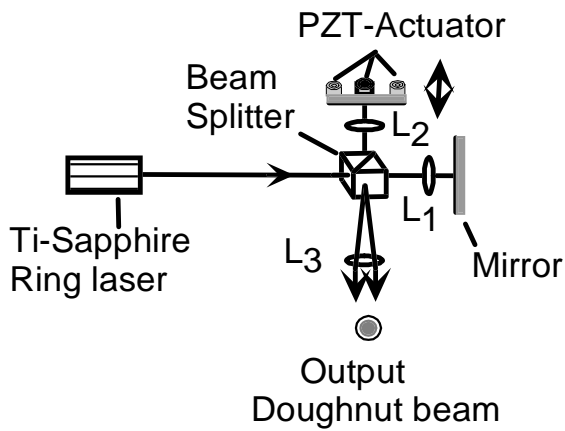 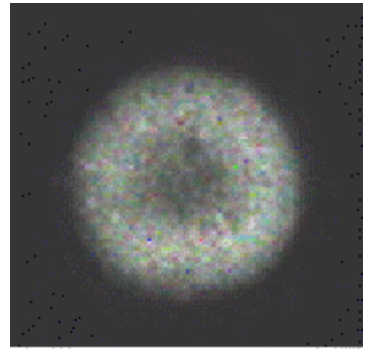

**Figure 1**: Doughnut beam generation setup     **Figure 2**: CCD image of doughnut beam